%% file: ms1054_xpaper_rev4.tex
\documentclass[useAMS,usenatbib]{mn2e}

\usepackage{epsfig}
\usepackage{psfig}
\usepackage{rotating}
\usepackage{rotate}
\input {epsf_mn.tex}

\def\bib{\parskip=0pt\par\noindent\hangindent\parindent
    \parskip=2ex plus .5ex minus .1ex}

\title[The AGN content of the $z=0.83$ cluster MS1054-0321]{The AGN
content of the z=0.83 cluster MS1054-0321}

\author[O. Johnson et al.]{O. Johnson\thanks{E-mail:
cocj@roe.ac.uk (OJ)}, P.~N. Best, O. Almaini\\
Institute for Astronomy, Royal Observatory Edinburgh, Blackford
Hill, Edinburgh, EH9 3HJ, UK\\}

\begin{document}

\date{Accepted  Received}

\pagerange{\pageref{firstpage}--\pageref{lastpage}} \pubyear{2002}

\maketitle

\label{firstpage}

\begin{abstract}

We present a survey of X-ray point sources in a 91 ksec Chandra ACIS-S
observation of the z=0.83 cluster MS1054-0321. We detect 47 X-ray
sources within the $8.3\arcmin\times8.3\arcmin$ field, of which two
are immediately confirmed from pre-existing spectroscopy to be
at the redshift of
the cluster.  At fluxes brighter than $S_{\rm(0.5 - 8 ~keV)} =
5\times10^{-15}$ erg s$^{-1}$ cm$^{-2}$ we find a $\sim 2\sigma$
excess compared to predictions from field surveys, consistent with an
excess of approximately $6$ AGN. If these sources are associated with
the cluster, they too are AGN with luminosities of order $L_{\rm(0.5 - 8
~keV)} \sim 10^{43}$ erg s$^{-1}$.  Combined with the identification
of 7 cluster AGN from deep radio observations (Best et al. 2002),
these observations suggest significantly enhanced AGN activity in
MS1054-03 compared to local galaxy clusters.  Interestingly, the
excess of X-ray detected AGN is found at radial distances of between 1
and 2 Mpc, suggesting they may be associated with infalling galaxies.
The radio AGN are seen within the inner Mpc of the cluster and are
largely undetected in the X-ray, suggesting they are either
intrinsically less luminous and/or heavily obscured.

\end{abstract}

\begin{keywords}

\end{keywords}

\section{Introduction}

Optical studies suggest that active nuclei are relatively rare in
cluster environments.  Surveying low redshift ($\overline{z}\sim0.04$)
cluster fields, Dressler et al. (1985) found only 1\% of cluster
members showed evidence for AGN
activity in their optical spectra.  A subsequent study of several distant
clusters revealed no increase in the cluster AGN fraction to
$z\sim0.5$ (Dressler et al. 1999).  Recently, however, surveys of
cluster fields at X-ray wavelengths are uncovering surprisingly large
numbers of point sources, many of which are confirmed to be cluster
AGN.
 
X-ray surveys of point sources within cluster fields have been
hindered by the high levels of soft X-ray flux from the hot
intra-cluster medium and by the traditionally poor angular resolution
of X-ray telescopes.  The current generation of X-ray observatories,
however, offer unprecedented resolution and positional accuracy, as
well as high sensitivity over the full X-ray band.  Furthermore,
multi-scale wavelet detection techniques (e.g. Freeman et al. 2000)
can now reliably separate small-scale point source emission from
surrounding larger-scale diffuse cluster emission.  Consequently,
several studies of point sources serendipitously observed in pointed
cluster observations have recently been published, nearly all of which
report an excess of point sources in these fields.

At low redshift, the excess of X-ray sources seems to be largely due
to low luminosity AGN (LLAGN) associated with the cluster.  Lazzati
et al. (1998), using a wavelet detection algorithm on ROSAT PSPC
data, examined the fields of A194 ($z=0.018$) and A1367 ($z=0.022$) and
found 26 and 28 sources, respectively, where only 9 were expected.
Sun and Murray (2002) reobserved A1367 with Chandra and resolved 59
point-like sources, 8 of which are confirmed cluster members.  The
identified sources, with typical X-ray luminosities
of a few $\times 10^{41}$ erg s$^{-1}$, have spectra consistent with
radiation from central nuclei, thermal halos, and stellar components.
In addition to these identified objects, a 2.5 $\sigma$ excess in
source counts at bright fluxes is found in both hard and soft bands.
If these unidentified sources are also associated with the cluster,
they have X-ray luminosities consistent with LLAGN and unusually low
optical luminosities.  A similar, though slightly more X-ray luminous
population was reported in A2256 ($z=0.06$) by Henry \& Briel (1991),
who found twice as many sources in a ROSAT PSPC observation than were
expected for a blank field.  Among these sources, they noted two
cluster member AGN with $L_{\rm (0.1-2.4 ~keV)} \sim 10^{42}$ erg
s$^{-1}$, as well as a number of optically `dark' objects with
comparable X-ray fluxes.

At higher redshifts, more luminous AGN are found in several clusters.
In the field of A2104 ($z=0.154$), Martini et al. (2002) find that all
X-ray sources with optical counterparts brighter than R=20 and B-R
colors consistent with the cluster sequence are indeed AGN in the
cluster, with $L_{\rm 2-10 ~keV}$ ranging from 10$^{41-45}$ erg s$^{-1}$.
Only one of the six has spectral characteristics which would have led
to it being optically identified as an AGN.  The AGN hosts in A2104 are
photometrically and spectroscopically similar to old, red galaxies,
but have a higher velocity dispersion than non-AGN cluster members,
suggesting some may be falling into the cluster for the first time.
Cappi et al. (2001) identify X-ray source excesses in 3C295 ($z=0.46$)
and RXJ0030 ($z=0.5$) at fluxes consistent with luminosities
at the cluster redshift of 10$^{42-43}$ erg s$^{-1}$, and both are 
found to contain
spectroscopically confirmed cluster AGN.  Finally, Pentericci et
al. (2002) report the association of 2, and possible association of as
many as 6, AGN with 
L$\sim10^{42-45}$ erg s$^{-1}$ in a
protocluster around the $z=2.16$ radio galaxy MRC 1138-262.

Not all moderate to high redshift clusters contain luminous AGN,
however.  Molnar et al. (2002) found an excess of sources around A1995
($z=0.32$) at fluxes consistent with starburst luminosities at the
cluster redshift.  They also surveyed MS0451 ($z = 0.55$) and found no
evidence for an excess of sources.  This is not inconsistent with
MS0451 housing a population similar to that seen in A1995, which would
be too faint to observe in the more distant cluster, but it does rule
out a population of more luminous sources such as those seen in 3C295
and RXJ003.  They conclude that not all clusters exhibit source
excesses at these flux levels, and suggest that different types of
sources appear to dominate in different clusters.

As part of a programme to investigate the prevalence of AGN activity
in galaxies as a function of redshift and dynamical state, we present
X-ray observations of one of the most distant known galaxy clusters.
At $z=0.83$, MS1054-03 is a well studied, rich (Abell class 3) cluster
with a bolometric 
$L_{X} = 1.2\times 10^{45} h^{-2}$ erg s$^{-1}$.  It
was the most distant cluster in the Einstein Medium Sensitivity Survey
X-ray selected cluster sample (Gioia et al. 1990) and has been the
subject of an extensive observing campaign yielding a Hubble Space
Telescope (HST) mosaic (van Dokkum et al. 2000), deep ground-based
infrared and optical imaging (Franx et al. in prep), and Keck 
spectroscopy (van Dokkum et al. 2000).
Substantial substructure in the distribution of galaxies is matched by
the diffuse soft X-ray emission, and weak-lensing analysis indicates
that the cluster is young, massive, and still relaxing (Hoekstra,
Franx, \& Kuijken 2000).  Though the Butcher-Oemler fraction,
$f_{B}=0.22\pm0.05$, is fairly low for this redshift, the fraction of
early-type galaxies in the central regions of the cluster (44\%) is
quite low and the fraction of merging systems (17\%) is quite high
(van Dokkum et al. 2000).  Extremely deep radio observations at 5 GHz have
revealed 6, and possibly 7, radio-loud AGN associated with the cluster (Best et
al. 2002; see \S\ref{radio}).  To probe the AGN content in the X-rays,
we analyze a 91 ksec Chandra exposure of the field obtained by 
Jeltema et al.(2001).

\S 2 describes the compilation of the X-ray sourcelist and the 
identification of optical counterparts.  In \S 3, the X-ray number 
counts in the MS1054-03 field are derived and the excess over the
expected counts in a `blank field' is identified.  Possible reasons
for the excess are discussed in \S 4, along with its relation to the 
radio source excess observed by Best et al. (2002).  Finally, we 
summarize our conclusions in \S 5.  Throughout the paper, we assume 
an $\Omega_{0}=0.3$, $\Lambda_{0}=0.7$, and H$_{0}$ = 70 km s$^{-1}$ 
Mpc$^{-1}$ cosmology.

\section{X-ray sourcelist}

\begin{figure}
\centering
\includegraphics[width=8.0truecm]{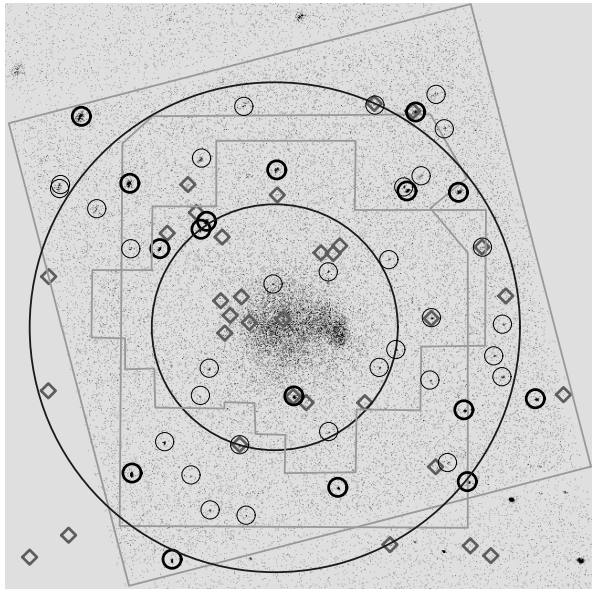}
\caption{X-ray point sources in the MS1054-03 field.  The
greyscale is binned full band ACIS-S data, displayed so that North is
up and East is to the left. The 47 detected
X-ray point sources within the ACIS-S3 chip are circled, with the 
darker circles indicating
0.5 - 8 keV flux greater than $\sim 5\times10^{-15}$ erg cm$^{-2}$
s$^{-1}$ (see \S 3).  The radio sources of Best et al. (2002) are marked
with diamonds.  The large circles are radii of 1 and 2 Mpc from the 
cluster center at 10 57 00.2, -3 37 36.0 (J2000). The outer grey box
delineates the edges of the S3 chip which is 8.3\arcmin on a side.
The irregular central contour is the FOV of HST imaging in this field.  
The roughly rectangular contour aligned North-South is the FOV over
which KECK spectroscopy is available for selected objects.  
\label{fig:fig1}}
\end{figure}

\begin{table*}
\begin{minipage}{165mm}
  \begin{tabular}{ccccccccc}
\hline
ID	&Name	&RA	&Dec	&Err	&Net	&Source	&0.5 - 8 keV Flux	&Hardness\\
	&(CXOU) &(J2000) &(J2000) &(arcsec) &Counts &Significance	&($10^{-14}$ erg cm$^{-2}$ s$^{-1}$) &Ratio\\

\hline
1 &105658.8-033850  &10:56:58.82  &-03:38:50.7  &0.50  &442.9  &111.7  &3.58$\pm$0.17	&-0.36$\pm$0.04	\\
2 &105710.4-034013  &10:57:10.49  &-03:40:13.9  &0.50  &302.8  &88.3  &2.46$\pm$0.14 	&-0.57$\pm$0.05	\\
3 &105655.6-034029  &10:56:55.65  &-03:40:29.4  &0.50  &273.1  &86.7  &2.18$\pm$0.13 	&-0.59$\pm$0.05	\\
4 &105646.3-034023  &10:56:46.31  &-03:40:23.3  &0.50  &251.4  &71.7  &2.14$\pm$0.13 	&-0.30$\pm$0.06	\\
5 &105646.5-033905  &10:56:46.56  &-03:39:05.7  &0.50  &199.4  &64.9  &1.64$\pm$0.11 	&-0.50$\pm$0.06	\\
6 &105650.6-033509  &10:56:50.66  &-03:35:09.1  &0.51  &163.0  &45.5  &1.31$\pm$0.10 	&-0.77$\pm$0.05	\\
7 &105710.6-033500  &10:57:10.65  &-03:35:00.9  &0.53  &176.8  &34.6  &1.45$\pm$0.11 	&-0.79$\pm$0.05	\\
8 &105705.1-033541  &10:57:05.11  &-03:35:41.9  &0.52  &125.2  &33.5  &1.00$\pm$0.09 	&-0.57$\pm$0.07	\\
9 &105708.4-033611  &10:57:08.48  &-03:36:11.5  &0.52  &115.5  &31.3  &0.92$\pm$0.08 	&-0.52$\pm$0.08	\\
10 &105646.9-033510  &10:56:46.95  &-03:35:10.6  &0.52  &112.7  &30.9  &0.91$\pm$0.08	&-0.93$\pm$0.03	\\
11 &105714.1-033348  &10:57:14.14  &-03:33:48.5  &0.56  &185.7  &30.4  &1.64$\pm$0.12	&-0.53$\pm$0.06	\\
12 &105705.5-033550  &10:57:05.52  &-03:35:50.5  &0.52  &84.9  &23.5  &0.67$\pm$0.07 	&-0.58$\pm$0.09	\\
13 &105700.0-033446  &10:57:00.05  &-03:34:46.4  &0.54  &69.9  &19.3  &0.58$\pm$0.07 	&-0.50$\pm$0.10	\\
14 &105641.4-033853  &10:56:41.40  &-03:38:53.7  &0.53  &59.0  &18.7  &0.49$\pm$0.06 	&-0.35$\pm$0.12	\\
15 &105650.0-033343  &10:56:50.03  &-03:33:43.8  &0.57  &85.2  &18.2  &0.72$\pm$0.08 	&-0.56$\pm$0.09	\\
16 &105707.6-034147  &10:57:07.61  &-03:41:47.1  &0.53  &48.7  &17.8  &0.78$\pm$0.11 	&-0.50$\pm$0.12	\\
17 &105708.1-033940  &10:57:08.14  &-03:39:40.0  &0.52  &48.4  &17.1  &0.40$\pm$0.06 	&-0.49$\pm$0.13	\\
18 &105648.8-033726  &10:56:48.90  &-03:37:26.0  &0.52  &48.6  &16.1  &0.39$\pm$0.05 	&-0.64$\pm$0.11	\\
19 &105702.7-033943  &10:57:02.73  &-03:39:43.4  &0.51  &36.5  &13.1  &0.28$\pm$0.04 	&0.08$\pm$0.17 	\\
20 &105652.6-033819  &10:56:52.67  &-03:38:19.8  &0.51  &36.4  &12.8  &0.29$\pm$0.05 	&-0.44$\pm$0.15	\\
21 &105643.8-033829  &10:56:43.81  &-03:38:29.8  &0.54  &32.5  &10.7  &0.27$\pm$0.04 	&0.10$\pm$0.17 	\\
22 &105705.4-033433  &10:57:05.48  &-03:34:33.6  &0.57  &33.9  &10.6  &0.28$\pm$0.05 	&-0.48$\pm$0.14	\\
23 &105650.8-033504  &10:56:50.90  &-03:35:04.8  &0.57  &22.5  &7.4  &0.18$\pm$0.04  	&-0.46$\pm$0.20	\\
24 &105713.0-033528  &10:57:13.04  &-03:35:28.3  &0.74  &27.9  &7.3  &0.23$\pm$0.04  	&-0.47$\pm$0.18	\\
25 &105715.6-033502\footnote{CXOU105715.7-033502 and CXOU105715.7-033506 are resolved by WAVDETECT as two separate detections in the full band, but are
detected as a single source in the soft band.}    &10:57:15.65  &-03:35:02.2  &0.60  &22.3  &7.2  &0.18$\pm$0.04	&-0.55$\pm$0.16\\
26 &105715.7-033506\Large{$^{a}$} &10:57:15.72  &-03:35:06.5  &0.63  &20.7  &6.4  &0.17$\pm$0.04	&-0.47$\pm$0.17\\
27 &105644.4-033807  &10:56:44.40  &-03:38:07.3  &0.54  &15.8  &5.8 &0.13$\pm$0.03   &0.32$\pm$0.24   	\\
28 &105651.4-033800  &10:56:51.46  &-03:38:00.5  &0.53  &15.1  &5.7  &0.12$\pm$0.03  &-0.77$\pm$0.16 	\\
29 &105704.9-033820  &10:57:04.92  &-03:38:20.7  &0.57  &17.5  &5.7  &0.14$\pm$0.03  &0.67$\pm$0.18  	\\
30 &105656.3-033636  &10:56:56.33  &-03:36:36.5  &0.54  &16.4  &5.3  &0.13$\pm$0.03  &-0.25$\pm$0.23 	\\
31 &105704.8-034053  &10:57:04.86  &-03:40:53.8  &0.56  &13.7  &5.0  &0.11$\pm$0.03  &0.11$\pm$0.26  	\\
32 &105702.4-033337  &10:57:02.42  &-03:33:37.8  &0.63  &14.7  &4.9  &0.12$\pm$0.03  &0.05$\pm$0.25  	\\
33 &105643.7-033733  &10:56:43.77  &-03:37:33.1  &0.58  &13.5  &4.8  &0.11$\pm$0.03  &-0.14$\pm$0.27 	\\
34 &105647.9-033401  &10:56:47.96  &-03:34:01.9  &0.66  &17.2  &4.8  &0.14$\pm$0.04  &0.42$\pm$0.23  	\\
35 &105649.0-033833  &10:56:49.00  &-03:38:33.5  &0.54  &10.6  &4.3  &0.08$\pm$0.02  &-0.55$\pm$0.25 	\\
36 &105647.7-034002  &10:56:47.73  &-03:40:02.6  &0.61  &12.3  &4.3  &0.10$\pm$0.03  &-0.17$\pm$0.29 	\\
37 &105710.5-033611  &10:57:10.58  &-03:36:11.6  &0.58  &10.4  &4.1  &0.08$\pm$0.02  &-0.20$\pm$0.31 	\\
38 &105648.5-033324  &10:56:48.56  &-03:33:24.8  &0.64  &12.3  &3.9  &0.10$\pm$0.03  &1.00$\pm$0.00  	\\
39 &105651.9-033623  &10:56:51.96  &-03:36:23.4  &0.60  &10.7  &3.9  &0.08$\pm$0.02 &-0.44$\pm$0.27 	\\
40 &105649.6-033452  &10:56:49.64  &-03:34:52.7  &0.59  &10.6  &3.8  &0.08$\pm$0.02  &-0.21$\pm$0.31 	\\
41 &105645.2-033609  &10:56:45.21  &-03:36:09.9  &0.62  &10.0  &3.8  &0.08$\pm$0.03  &0.17$\pm$0.30  	\\
42 &105700.3-033649  &10:57:00.32  &-03:36:49.6  &0.55  &12.9  &3.7  &0.10$\pm$0.03   &0.52$\pm$0.25  	\\
43 &105702.2-034059  &10:57:02.26  &-03:40:59.5  &0.53  &8.8  &3.6  &0.07$\pm$0.02   &-0.41$\pm$0.30 	\\
44 &105652.9-033336  &10:56:52.97  &-03:33:36.3  &0.68  &10.2  &3.3  &0.08$\pm$0.03  &-0.33$\pm$0.30 	\\
45 &105705.5-033850  &10:57:05.56  &-03:38:50.0  &0.60  &8.3  &3.2  &0.06$\pm$0.02   &0.64$\pm$0.26  	\\
46 &105706.2-034016  &10:57:06.24  &-03:40:16.0  &0.59  &7.6  &3.1  &0.06$\pm$0.02   &0.85$\pm$0.18  	\\
47 &105656.3-033929  &10:56:56.30  &-03:39:29.1  &0.54  &7.5  &3.0  &0.06$\pm$0.02   &0.17$\pm$0.35  	\\
\hline
\hline
  \end{tabular}
\caption{X-ray point sources detected in the MS1054-03 field.  The quoted
error is the average of the positional error in RA and DEC added in
quadrature to the mean astrometric error.  The source significance is
the output value from WAVDETECT, defined in \S 3, and equal to the
source counts divided by the Gehrels error on the background counts.
The flux values and associated Poisson errors were calculated assuming
$\alpha = 0.7$ powerlaw spectra.     
\label{tab:tab1}}
\end{minipage}
\end{table*}

Our analysis uses an archival 91 ksec exposure of MS1054-03 taken with
the back-illuminated Chandra ACIS-S3 detector on 2000 April 21-22, and
first published in Jeltema et al. (2001).  We reduced the data with standard
{\small CIAO} tools, using {\small CALDB} version 2.1 which includes an improved ACIS-S3
gain map.  There were three periods of significant background flaring
during the integration which we excluded, leaving a useable exposure
of 74 ksec. The data were filtered by pulse height energy into soft
(0.5 - 2 keV), hard (2 - 8 keV), and full (0.5 - 8 keV) bands.
Exposure maps were made in each band to estimate variation in
effective exposure across the sky due to instrumental effects and
source dither.  These maps are energy dependent, and were made
assuming a power-law spectrum with a photon index $\gamma$ = 1.7, 
typical of unobscured AGN at our flux limit ($\sim 10^{-15}$ erg s$^{-1}$
cm$^{-2}$; see Figure 3 of Tozzi et al. 2001).
The {\small WAVDETECT} source detection algorithm 
(Freeman et al. 2000) was used to
construct independent point source lists in each band.  A significance
threshold of $6.1 \times 10^{-7}$ was used, corresponding to an average
detection of less than one false source over the area of a single,
full-resolution ACIS chip.  We further limited our source lists by
rejecting sources with a {\small WAVDETECT} source significance 
parameter of less than 3.0.  Though this cut-off is arbitrary - the 
few sources with
lower values may be real - visual inspection suggests it is a
reasonable, conservative limit and its definition in terms of this
parameter is computationally convenient.  We find 47 sources in the
full band, 33 in the soft band, and 24 in the hard band.  Excepting a
soft band only detection of a peak in the cluster emission, there are
no soft or hard band detections which are not also full band
detections.  We will use only the full band source list in the rest of
this analysis, and fluxes and luminosities are quoted for the 
observed 0.5 - 8 keV band.    

We used the Starlink {\small ASTROM} package to astrometrically correct the
X-ray source positions.  AGN with accurately known coordinates in the
optical and/or radio were aligned with their Chandra counterparts in a
6-parameter fit.  Eleven sources were used in total, and achieved a good
coverage of the chip.  The RMS fit residuals were all $<$ 1.2\arcsec,
with a mean of 0.48\arcsec.  The positional errors include the error
in RA and Dec from {\small WAVDETECT}, of order 0.25\arcsec for a
typical source, with a 0.5\arcsec astrometric error added in quadrature. 

Hardness ratios
for each source are defined as $\frac{H-S}{H+S}$, where $H$ and $S$
are the counts in the 2 - 8 keV and 0.5 - 2 keV bands, respectively. 
 Background subtracted hard
and soft band counts were obtained through aperature 
photometry in the {\small WAVDETECT} source regions.
The detected sources are listed in Table \ref{tab:tab1}, and
shown in Fig. \ref{fig:fig1}.

\subsection{Optical counterparts}
\label{opt}

As illustrated in Figure \ref{fig:fig1}, there is a wealth of optical
observations of MS1054-03, none of which completely match the ACIS-S3
$8.3\arcmin\times8.3\arcmin$ field of view (FOV).  Deep HST mosaics made with the
Wide-field Planetary Camera 2 (WFPC2) in the F606W and F814W filters
cover a central area of $\sim 5\arcmin\times5\arcmin$, corresponding to a
radius of $\sim$ 1 Mpc from the cluster center at the cluster redshift
(van Dokkum et al. 2000).  Spectroscopy of selected objects has been
obtained with the Keck telescope over an area 
of $\sim 6.8\arcmin\times6.8\arcmin$ 
yielding redshifts for over 200 objects
in the field, 130 of which are confirmed cluster members (van Dokkum
et al. 2000). 

Where coverage allowed, optical counterparts for X-ray sources were
identified from the Keck data and the HST mosaic.  Each identification 
was examined by eye, and the largest
observed offset was 1.35 arcsec.  All 15 X-ray sources in the
HST FOV were identified.  19 out of 31 sources in the Keck FOV were
identified, of which 7 were in the HST FOV.  In all, optical
counterparts were found for 27 of the 47 X-ray sources.  Restframe U -
B colors are available for only 11 sources within the HST FOV (cf. van Dokkum
et al. 2000); none of these lie within 0.2 mags of the cluster sequence.  
However, as the HST data covers only the inner regions of the cluster, 
available optical color data can give no indication of the likely 
cluster membership of sources further from the cluster center 
(see \S\ref{excess}, \S\ref{radial}).

Spectroscopic redshifts were available for just 7 of the X-ray
sources, of which two were found to be at the cluster redshift.  
CXOU105702.7-033943 (source 19) is a visually
confirmed match to KECK ID 564.  It has extended morphology, is on
the cluster sequence, and is also detected in the radio (see \S
\ref{radio}).  It has a rest frame $L_{\rm(0.9-14.6 keV)} = 5\times
10^{42}$ erg
s$^{-1}$, and a hardness ratio of 0.08 which corresponds to an
obscuring column of a few times $10^{22}$ atoms cm$^{-3}$ for a
typical powerlaw spectrum with energy index, $\alpha$, of 0.7 to 1.  
CXOU105710.6-033500 (source 7) is consistent with being a match 
to KECK ID 2152, which is brighter and bluer than the cluster
sequence.  It has a rest frame $L_{\rm(0.9-14.6 keV)} = 2.6\times10^{43}$ erg
s$^{-1}$, and a hardness ratio of -0.79, suggesting a somewhat
softer spectrum with $\alpha\sim1.6$ for an unobscured powerlaw. 

\section{Source counts}

We are interested in how many more of the detected X-ray sources in the MS
1054 field may be associated with the cluster.  In the absence of complete
spectroscopic identifications, we begin by looking for an excess over
the source counts expected in a cluster-free field.

\subsection{Determining the source counts}

Determination of the survey area available to this observation at a
given flux is complicated by variation of the effective exposure over
the field due to instrumental effects, and by the diffuse emission
from the hot cluster gas.  We account for these effects by calculating
a flux limit map, which indicates the flux of the faintest source
which would have been included in our source list at each position on
the chip.  The sky area available at a given flux limit is then the
summed value of all pixels with lower values.

We have chosen to limit our catalog in terms of the {\small WAVDETECT} source
significance parameter, $\sigma_{src}$, defined as

\[ \sigma_{src} = \frac{C_{src}}{1 + \sqrt{0.75 + C_{bkg}}} \]

\noindent where $C_{src}$ are the net source counts and $C_{bkg}$ are the
background counts within the source region (Chandra X-ray Center, 2002).

The detected source counts are related to the incident source flux,
$S$ (erg cm$^{-2}$ s$^{-1}$), the effective exposure area, EA
(cm$^{2}$), and the exposure time, $\tau$ (s), by

\[ C = S \times EA \times \tau \times K \]
  
\noindent where K is a conversion factor from ergs to counts assuming an $\gamma
= 1.7$ powerlaw spectrum.

The background counts in the source region depend on the local
background level, $B$, as well as the on size of the source, $D$,
which for point sources mainly reflects the off-axis degradation of
the Chandra PSF.  We treat the diffuse cluster emission as local background.

Therefore the flux limit, $S_{lim}$, for inclusion in our $\sigma
_{src} > 3$ source list is

\[S_{lim} = 3 \times \frac{1 + \sqrt{0.75 + (B \times D)}}{EA \times
\tau \times K} \]

In practice, maps of the same size and resolution as the data
were made for quantities $B$, $D$, and $EA$, allowing calculation of
$S_{lim}$ for each pixel.  The derivation of the exposure map is
discussed in the previous section.

To produce a map of the true background plus the diffuse cluster
emission, we first subtracted all identified point sources using
the {\small CIAO} tool {\bf dmcalc}.  We filled the resulting
``holes'' with pixel values sampled from the Poisson distribution
whose mean and standard deviation equalled that of the 
surrounding background pixels with
{\bf dmfilth}.  This source-free image was then smoothed with a
30\arcsec ~Gaussian kernal using {\bf csmooth}.  Our background
estimate was compared to the output background image from {\small
WAVDETECT} to ensure no significant error was introduced by the
inclusion of low significance ``sources'' in our map which are
excluded during the wavelet detection process.  The {\small WAVDETECT}
background image was not used directly due to minor artifacts left
from the iterative subtraction of bright sources.

In theory, the size of a detected point source should equal the size
of the Chandra PSF, which depends largely on source angular distance
from the optical axis, and also to some extent on source energy.  The
Chandra PSF library contains radially averaged encircled energy
curves for several off-axis angles at a number of calibration
energies, and these data were interpolated to create a map of PSF
size over the extent of the S3 chip.  We used the calibration data at
1.49 keV, as this was the value chosen for the {\small WAVDETECT} input
parameter {\tt eenergy} which specifies the PSF curves used in the
detection algorithm.  Comparing the source sizes reported by {\small WAVDETECT}
for detected sources with those predicted by our source size map, we
noted the {\small WAVDETECT} values were significantly higher at low off-axis
angles.  This appears to be due to a trend toward higher values of the
WAVDETECT source extent parameter, {\small PSFRATIO}, near the optical
axis, and not to an inconsistency in the expected PSF size (WAVDETECT
{\small PSF\_SIZE} ).  Since our estimation of the limiting flux hinges
on whether the source would have been detected by the wavelet
detection algorithm, we have empirically fit the dependence of
WAVDETECT source size on off-axis angle and included this in our
source size map.

\subsection{The source excess}
\label{excess}

\begin{figure}
\centering
\hspace{-8mm}
\includegraphics[width=65mm,angle=90]{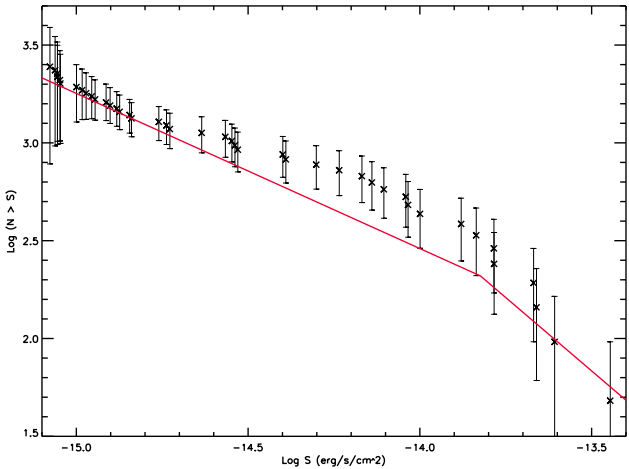}
\caption{Log N - Log S for MS1054-03 field.  Points are number counts
for the cluster field, while the solid line represents the counts expected 
from a blank field.  At fluxes brighter than $1.5\times10^{-14}$
erg s$^{-1}$ cm$^{-2}$, the slope is Euclidean; the slope at fainter fluxes
is the best fit to the EXDS counts from Manners et al. (2002).      
\label{fig:logn}}
\end{figure}

Figure \ref{fig:logn} shows the resulting number counts for the MS
1054 field.  We have overplotted the best fit to number counts from
the ELAIS Deep X-ray Survey (EXDS) presented in Manners et al. (2002).  
We choose the EXDS number counts computed in the 0.5 - 8 keV band for
ease of comparison.
Comparisons of the EXDS counts in soft and hard bands show them to be
fully consistent with counts derived from other Chandra deep surveys,
notably that of Mushotsky et al. (2000) obtained with the S3 chip.  A clear
excess of sources is seen at moderate fluxes in the MS1054-03 field, 
reaching $\sim$ 50\% at
$5\times10^{-15}$ erg cm$^{-2}$ s$^{-1}$, which corresponds to a
source luminosity of $\sim 10^{43}$ erg s$^{-1}$ at the redshift of the
cluster.  Sources with fluxes in the range contributing to this excess 
are highlighted in Figure \ref{fig:fig1}, where they are seen to avoid
the cluster center.  

We have repeated the analysis with sources detected independently in
the soft and hard bands.  Consistent with earlier papers (Cappi et
al. 2001, Pentericci et al. 2002), we detect an excess of comparable
significance in the soft band sources but no significant excess of
hard sources.

\begin{figure}
\centering
\hspace{-8mm}
\includegraphics[width=65mm,angle=90]{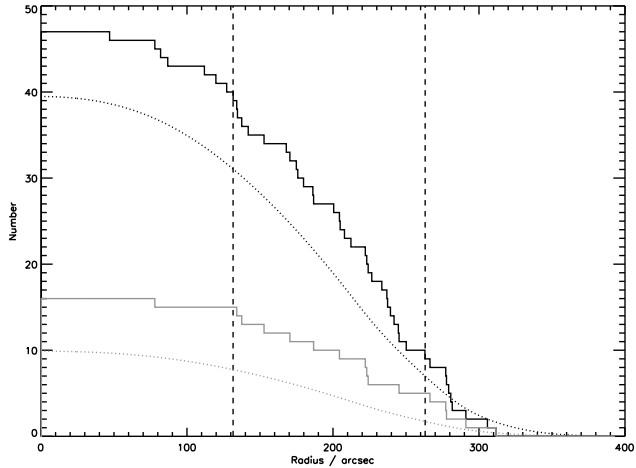}
\caption{Cumulative radial histogram of sources in MS1054-03 field,
moving in from
a distance of 400\arcsec ~to the cluster center at 10 57 00.2, -3 37
36.0 (J2000).  The detected sources (solid histogram) are compared with 
sources expected from a cluster-free field (dotted curve; see text).  The upper, dark curves are
for the full X-ray sample, and the lower, light curves are for sources
with  S $> 5\times10^{-15}$ erg s$^{-1}$ cm$^{-2}$.
The vertical lines indicate radii of 1 and 2 Mpc
at the cluster redshift. 
\label{fig:cumhist}}
\end{figure}

To investigate where in the field these excess sources are, we compare
radial source histograms for the data and a simulated 
field with the same spatially-dependent sensitivity.  The latter was
made by converting the flux limit map described above to an ``expected
source'' map, using the EXDS number counts to obtain the fractional
expected source count per pixel at each limiting flux.  Cumulative
radial source counts for our data and for the cluster-free field are
compared in Figure \ref{fig:cumhist} from a distance of 400\arcsec ~to
the cluster center.  We show both the sources detected to the survey
limit (upper, dark line), and those contributing to the excess at fluxes
above $5\times10^{-15}$ erg s$^{-1}$ cm$^{-2}$ (lower, light line).  The 47
sources detected at all fluxes compare with 40 sources expected over
the same area for a cluster-free observation.  As sources
105715.7-033502 and 105715.7-033506 may constitute a single object
(see footnote to Table 1), we reliably detect 6 sources more
than we would expect in a blank field.  This is only a 1$\sigma$
excess overall, and is fully consistent with observed levels of cosmic
variance in Chandra deep survey fields (Manners et al. 2002).  At
fluxes higher than $5\times10^{-15}$ erg s$^{-1}$
cm$^{-2}$, however, we expect only 10 sources, and detect 16, which
corresponds to an excess of $\sim 2\sigma$.  It is clear from Figure
\ref{fig:cumhist} that this excess arises in the outer regions of the
cluster between 1 and 2 Mpc.  The two confirmed X-ray detected cluster AGN, one brighter and one fainter than the $5\times10^{-15}$ erg s$^{-1}$ limit, are found 
134\arcsec ~and 222\arcsec ~from the cluster center, corresponding to distances of $\sim$
1.0 and 1.7 Mpc.

\section{Discussion}

\subsection{Cosmic variance}
While the probability of randomly observing as many excess sources as
are seen in this field over the entire flux range is 0.19, the result
at brighter fluxes has a probability of only 0.05.  As noted by Cappi
et al. (2001), the magnitude of the excess -- $\sim$ 50\% at $5\times
10^{-15}$ erg s$^{-1}$ cm$^{-2}$ -- is higher than the 20\%-30\%
variation seen in deep field surveys due to cosmic variance.
Furthermore, the fact that this magnitude of excess sources has been
detected in all but one of the seven Chandra pointed observations of
clusters argues strongly against explanation by cosmic variance.

\subsection{Gravitational lensing}

As noted in Cappi et al. (2001), gravitational lensing of background 
sources by the cluster is also not a viable explanation for the 
observed effect.  An intervening lens produces two
effects which alter the apparent source counts in opposite senses.
The apparent luminosity of sources is boosted by the lens, but the sky
area in the lensed regions is decreased.  Whether an increase or a
deficit of sources is observed depends critically on the slope of the
number counts being lensed.  If the unlensed integral source counts
are described as a powerlaw

\[ N(>S_{\nu}) \propto S_{\nu}^{-\alpha} \]

\noindent then the counts boosted by a factor $\mu(\theta)$ will be

\[N'( >S_{\nu}) = \frac{1}{\mu(\theta)} N(>\frac{S_{\nu}}{\mu(\theta)}) = \mu^{\alpha-1}N(>S_{\nu}) \]

\noindent where $\theta$ is the angular distance from the lens center.
For source populations with an integral slope $\alpha < 1$ the
apparent source counts will be depleted, while those with steeper
slopes will be enhanced. 

The effect of lensing by clusters of galaxies on the resolved and unresolved 
components of the X-ray background (XRB) is discussed at length in
Refregier \& Loeb (1997), who predict an overall depletion of resolved
sources due to the shallow slope of the XRB number counts at fluxes
fainter than $\sim 10^{-14}$ erg s$^{-1}$.  The measured slope
of the full band EXDS number counts over the flux range 
(1.1 - 15)$\times 10^{-15}$ erg s$^{-1}$cm$^{-2}$ is $\alpha \sim
0.8$, steepening to a Euclidean slope at brighter fluxes 
(Manners et al. 2002).  Lensing in the MS1054-03 field should lead to 
a deficit rather than an excess of in the number counts at all fluxes
fainter than the break.   

To investigate whether this effect might be significant enough to
significantly deplete the source counts in the inner regions of the
cluster, we model the cluster as a single isothermal sphere (SIS) with
a central density profile $\rho(r)\propto r^{-2}$.  While clearly a
gross approximation to the complex mass structure observed in MS1054-03,
the SIS profile provides a simple estimate of the magnitude of the
lensing effect.  For a SIS lensing potential, the lensing factor,
$\mu(\theta)$, can be expressed as

\[\mu(\theta) = \left | 1 - \frac{\theta_{E}}{\theta} \right | ^{-1} \]

\noindent where $\theta_{(E)}$ is the Einstein radius.  Following
Blandford \& Narayan (1992), $\theta_{(E)}$ is given in terms of the
angular diameter distances between the source and the lens, $D_{LS}$,
and between the source and the observer, $D_{OS}$:

\[ \theta_{E} = \frac{4\pi\theta^{2}}{c^{2}}\frac{D_{LS}}{D_{OS}} = 2.6\arcsec \sigma_{300}^{2}\frac{D_{LS}}{D_{OS}}\]

\noindent where $\sigma_{300}$ is the velocity dispersion of the lens,
in units of 300 km s$^{-1}$.

The velocity dispersion of MS1054-03 has been measured dynamically (Tran
et al. 1999; $\sigma \sim 1170 \pm 150$ km s$^{-1}$) and estimated
through weak lensing analysis (Hoekstra et al. 2000; $\sigma =
1215^{+63}_{-67}$ km s$^{-1}$).  We adopt a value of 1200 km
s$^{-1}$.  

The effect of lensing on the measured source counts is less
than 5\% beyond $\sim$ 1Mpc and less than 10\% beyond $\sim$ 0.5 Mpc.
As we expect detection of only a few sources interior to 0.5 Mpc in a
blank field (see Figure \ref{fig:cumhist}), the lensing deficit is
effectively negligible at all radii.

\subsection{Radio overlap}
\label{radio}

Best et al. (2002) conducted extremely deep radio observations of
MS1054-03 at 5 GHz with a full-width-half-power primary antenna beam
diameter of $9\arcmin$.  They detected 34 sources to a 6$\sigma$ level of 32
$\mu$Jy, compared with 25 expected from a blank field.  The excess
radio source counts are all found within 1 Mpc of the cluster center,
and 8 sources are spectroscopically confirmed cluster members.  
Of the
cluster radio sources, seven appear to be powered by AGN.
On the basis of radio flux density to optical emission line flux
ratio, Best et al. (2002) report six cluster AGN and one ambiguous source
with a ratio between that expected for AGN and that expected for
star-forming galaxies.  The ambiguous source is detected in the X-ray
image with a $L_{\rm(0.5-8 ~keV)}\sim 10^{43}$ erg s$^{-1}$, however, which 
confirms it is an AGN.  
Four of these host galaxies are in merging systems,
suggesting the activity may be induced by interactions.

Six sources are detected in both radio and X-ray, only one of which
is at the cluster redshift.  The single cluster AGN detected in both
the X-ray and the radio is CXOU105702.7-033943 (source 19; Best et
al. source 14).  As stated earlier, the optical counterpart has
galactic morphology and colors on the cluster sequence and the X-ray
hardness ratio is consistent with a typical AGN powerlaw spectrum with
moderate ($\sim 10^{22}$ atoms cm$^{-3}$)  intrinsic absorption.  
The radio and X-ray luminosities are in
good agreement with the Elvis et al. mean SED for radio-loud AGN (see below,
and Figure \ref{fig:radcomp}).

The fraction of X-ray sources detected in the radio -- 6 out of 47 --
is consistent with the observation that $\sim$ 10\% of all AGN are
radio-loud (e.g. Sramek \& Weedman 1980).  If this fraction holds
within the cluster, we would expect no more than one of the two
confirmed or six suspected X-ray AGN to be radio-loud.  Referencing
the mean SED of Elvis et al. (1994), the expected radio flux for a
radio-quiet QSO with $L_{\rm{(0.5-8 ~keV)}}\sim10^{43}$ erg s$^{-1}$
cm$^{-2}$ is an order of magnitude lower than the radio detection
limit.  The detection of just one of the X-ray AGN in MS1054-03 at
radio wavelengths is therefore consistent with expectation.

\begin{figure}
\begin{center}
\hspace{-5mm}
\includegraphics[width=65mm,angle=90]{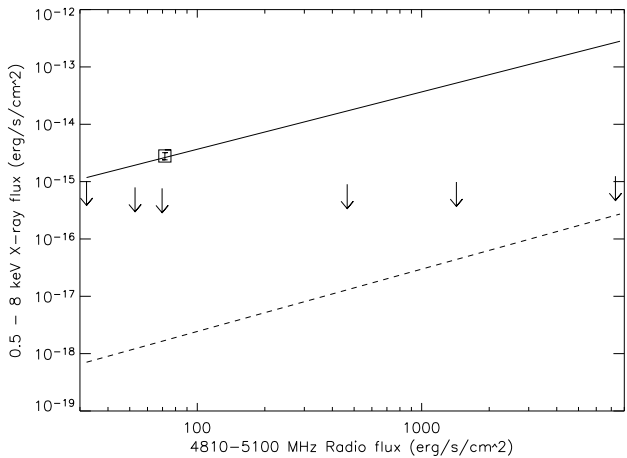}
\caption{Radio flux versus expected X-ray flux for the 7 confirmed
cluster radio sources.  The upper, solid line makes the X-ray flux predicted for
the measured radio flux by the Elvis et al. (1994) mean radio-loud QSO
SED; the lower, dotted line marks the flux predicted by the Sambruna \& Eracleous
(1999) relation for radio-loud AGN.  (See text for details.)  The
upper limits indicate
the limiting flux of the X-ray image at the location of the radio
sources, and the square shows the measured X-ray flux of CXOU105702.7-033943.     }
\label{fig:radcomp}
\end{center}
\end{figure}

The small fraction of radio sources detected in the X-ray -- 6 out of 34 --
is perhaps more surprising.  While many of the field radio sources are
likely to be starburst galaxies which we would not expect to detect in
X-ray observations at this depth, at least 7 of the sources are
radio-loud cluster AGN which should emit strongly in the
X-ray.  However, since the radio to X-ray flux ratio in AGN is seen to
vary by several orders of magnitude (Brinkmann et al. 2000, Bauer et
al. 2002), it is difficult to estimate the expected X-ray luminosity
of the radio AGN.
In figure \ref{fig:radcomp} we plot the radio flux of the seven
cluster radio AGN versus the expected X-ray flux calculated
from the Elvis et al. (1994) mean radio-loud QSO SED.  Following the Elvis
et al. SED, all seven sources could have been detected in our X-ray
observation.  Though scatter around the mean might account for the
non-detection of the fainter radio sources, the
brightest sources have predicted X-ray fluxes which are 1 to 2.5
orders of magnitude brighter than the flux limit.  

Dense absorbing columns could provide a possible explanation.
Obscuration of 12 -- 70\% of the 0.5 - 8 keV X-ray flux expected
from the faintest four radio detections would be sufficient to 
put them below the X-ray detection limit.  This level of absorption
would result from relatively modest columns of $\sim 5\times10^{21}$
atoms cm$^{-3}$ up to a few times $10^{23}$ atoms cm$^{-3}$.  For the
brightest three radio sources, 95\% to 99\% of the expected X-ray flux 
would need to be absorbed to push them below the detection limit.
Absorbing columns of $10^{24-25}$ atoms cm$^{-3}$, 
as are found to be relatively common in Seyfert 2s (e.g. Maoilino et al. 1998), would be needed for
these three objects.

It is possible the Elvis et al. mean SED, derived from a sample of 18
luminous X-ray selected radio-loud QSOs, is not representative of the
weaker radio AGN populating MS1054-03.  Sambruna \& Eracleous (1999) include
weaker radio galaxies in their X-ray survey which, in the unified
model, contain substantial absorption along the line of sight and may
more closely resemble the sources in our cluster.  The dotted line in
Figure \ref{fig:radcomp} shows their empirical relation of lobe 5
GHz luminosity to hard X-ray flux, assuming that lobe emission
dominates the radio flux, and extrapolating hand band X-ray flux to
full band flux using an unobscured powerlaw spectrum with an energy
index, $\alpha =0.7$.  The resulting expected X-ray fluxes are likely
too high, as non-negligible radio flux may arise from the compact core
and obscuring material having little effect on the hard band flux may
significantly absorb flux at lower energies.  However, even these
optimistically high fluxes suggest that all seven radio AGN in
MS1054-03 may be too faint to be observed in the X-ray.

A third possibility is that the radio luminosities may be contaminated
by a stellar component.  The low optical emission line flux to radio
flux density values, as well as the red optical colours of the radio
sources which are consistent with the cluster sequence, argue against
a post-starburst origin.  However, we cannot entirely rule out some
amount of the radio emission arising from stellar activity rather than
AGN activity, and this could help to explain the X-ray non-detections.
  
If the stellar contamination to the radio fluxes is low, and if the
usual radio-loud to radio-quiet ratio holds throughout the cluster,
the radio-detected AGN may trace a larger AGN population nearer to the
the cluster core.  For the 7 detected radio-loud AGN, one would expect
$\sim 70$ radio-quiet counterparts.  Like the radio-loud but X-ray quiet
sources, these would need to be either heavily obscured or simply much
less luminous than the detected X-ray sources to avoid detection in
the current data.  While a vast increase in the fraction of 
heavily obscured objects in the inner regions of the cluster seems
improbable, a population of less luminous AGN, such as those observed in 
clusters at low redshift, cannot be ruled out.  Alternatively, the
radio-loud to radio quiet fraction within the core of MS1054 may be 
elevated due to a dependence on AGN luminosity and/or environment.

\subsection{Radial distributions}  
\label{radial}

The X-ray detected AGN are not distributed randomly within the cluster
but tend to populate the outer 1-2 Mpc, suggesting they may just be
falling into the cluster.  Similarly, Martini et al. (2002) note that
while the galaxies hosting the cluster AGN in A2104 have colors on the
cluster sequence, their velocity dispersion is larger than the mean
cluster dispersion, perhaps suggesting recent infall.  If galaxies
hosting luminous cluster AGN are found to be recent arrivals to the
dense cluster environment, it may indicate the AGN activity is induced
by infall.  It is an open debate whether introduction into a cluster
environment serves purely to quench star formation activity in
infalling field galaxies, or results in an initial enhancement in
activity before quenching begins (cf. Poggianti 2002, and references
therein).  The post-starburst `k+a' galaxies seen in abundance in
distant clusters (Dressler et al. 1999) may be evidence of the 
latter scenario, if it is
found that starburst activity in the field is not sufficient to
account for them.  If the conditions prompting starbursts in infalling
members also serve to fuel AGN, the detection of an excess of AGN at
the edges of dynamically active clusters may be expected, and could be
a discriminator between the two scenarios described above.

Finally, it is worth emphasizing in this context that the AGN
identified by radio observations are found somewhat nearer to the
cluster center ($< 1$ Mpc).  If they are the radio-loud
members of a population of weaker AGN which are below 
the X-ray flux limit, they may indicate a decrease of AGN activity in
host galaxies which have been longer resident in the dense cluster environment.

\section{Conclusions}

We have detected an excess of point sources in the field of the rich
cluster MS1054-03 at $z = 0.83$.  The excess is most significant at
fluxes brighter than S$_{\rm (0.5 -8 ~keV)} = 5 \times 10^{-15}$ erg
s$^{-1}$ cm$^{-2}$ (corresponding to a rest frame $L_{(0.9-14.6 keV)} \simeq
10^{43}$ erg s$^{-1}$ for cluster members) and arises in the 
outer parts of the cluster
field, at angular separations consistent with radial distances of 1 -
2 Mpc from the cluster center.  The observed source counts cannot be
accounted for by cosmic variance, and gravitational lensing of
background counts by the cluster is expected to be negligible across
the field.  We conclude the most likely explanation of the excess
counts are a number of luminous (L $\sim 10^{43}$ erg s$^{-1}$) AGN
associated with the cluster.  The X-ray excess arises between 1 and 2
Mpc from the cluster center, exterior to the excess of radio sources
identified by Best et al. (2002).  Follow-up observations will be
necessary to completely identify the X-ray detections in this field,
but the association of two luminous AGN with the cluster is already
confirmed from existing observations.  CXOU105710.6-033500 (source 7)
is 1.7 Mpc from the cluster center with a luminosity of $2.6\times
10^{43}$ erg s$^{-1}$, and CXOU105702.7-033943 (source 19) is 1.0 Mpc
from the cluster center with a luminosity of $5.0\times 10^{42}$ erg
s$^{-1}$.  Including the further 6 radio AGN detected at 5 GHz by Best et
al. (2002), there are at least 8 luminous active nuclei associated
with MS1054-03.

This study augments earlier results, confirming that luminous AGN can
and do reside in rich cluster environments.  The luminosities of $\sim
10^{43}$ erg s$^{-1}$ we observe at $z = 0.83$ suggest this population
is similar to that seen in 3C295 and RXJ0030 at $z \sim 0.5$ (Cappi et
al. 2001).  Fainter populations such as the LLAGN seen in nearby
clusters and the starburst population seen in A1995 at $z=0.32$ may
also be present in higher redshift clusters such as MS1054-03.  The
radio AGN detected by Best et al. (2002) but not detected in the X-ray
may be evidence of just such a weaker population in MS1054-03.
However, there is no low redshift equivalent to the population of
luminous sources which are apparently common in high redshift
clusters like MS1054-03.  As not all clusters exhibit 
a population of point sources,
the extent and strength of AGN activity within a cluster is likely
dependent on both redshift and dynamical state.  Further study will be
directed to examining these relations.  Finally, the fact that the
luminous X-ray sources in MS1054-03 are observed only at the edges of
the cluster is suggestive of infall induced activity, the quenching of
which may give rise to fainter AGN at smaller cluster-centric radii.
Exploring this possibility is another motivation for further complete
surveys of AGN in distant clusters.

\section*{Acknowledgments}

We thank Pieter van Dokkum for his generous help with this work, and
the referee for helpful comments. 
OJ thanks Ignas Snellan and Andy Lawrence for useful discussions, and
the School of Physics, University of Edinburgh for partial funding.  
PNB and OA gratefully acknowledge the support of the Royal Society.

\section*{References}

\bib Bauer F.~E., Alexander D.~M., Brandt W.~N., Hornschemeier A.~E., 
Vignali C., Garmire G.~P., Schneider D.~P., 2002, AJ, 124, 2351

\bib Best P.~N., van Dokkum P.~G., Franx M., Rottgering H.~J.~A., 2002, MNRAS, 330, 17

\bib Blandford R.~D. \& Narayan R., 1992, ARA\&A, 30, 311

\bib Brinkmann W. et al., 2000, A\&A, 356, 445 

\bib Cappi M. et al., 2001, ApJ, 548, 624

\bib Chandra X-ray Center, 2002, The CIAO Detect Manual (CIAO Software
Release V2.2.1), (Cambridge: Chandra X-Ray Center)

\bib Dressler A., Thompson I.~B. Shectman S.~A., 1985, ApJ, 288, 481

\bib Dressler A., Smail I., Poggianti B.~M., Butcher H., Couch W.~J.,
Ellis R.~S., Oemler A., 1999, ApJS, 122, 51

\bib Franx M. et al., in preparation 

\bib Freeman P.~E., Kashyap V., Rosner R., Lamb D.~Q., 2002, ApJ, 138, 185

\bib Gioia I.~M., Maccacaro T., Schild R.~E., Wolter A., Stocke J.~T.,
Morris S.~L., Henry J.~P., 1990, ApJS, 72, 567

\bib Henry J.~P. \& Briel U.~G., 1991, A\&A, 246, L14

\bib Hoekstra H., Franx M., Kuijken K., 2000, 532, 88 

\bib Jeltema T.~E., Canizares C.~R., Bautz M.~W., Malm M.~R., 2001, ApJ, 562, 124

\bib Lazzati D., Campana S., Rosati P., Chincarini G., Giacconi R.,
1998, A\&A, 331, 41

\bib Manners J. et al., 2002, MNRAS, in press

\bib Martini P., Kelson D., Mulchaey J.~S., Trager S.~C., 2002, ApJ,
576, L109

\bib Molnar S.~M., Hughes J.~P., Donahue M., Joy M., 2002, ApJ, 573,
L91

\bib Mushotsky R.~F, Cowie L.~L, Barger A.~J., Arnaud K.~A, 2000,
Nature, 404, 459

\bib Pentericci L., Kurk J.~D., Carilli C.~L., Harris D.~E., Miley
G.~K., Rottgering H.~J.~A., 2002, A\&A, in press (astro-ph 0209392)

\bib Poggianti B. in "Galaxy Evolution in Groups and Clusters", 2002, 
eds. Lobo, Serote-Roos and Biviano, Kluwer, in press (astro-ph 0210233)

\bib Refregier A. and Loeb A., 1997, ApJ, 478, 476

\bib Sun M., Murray S.~S., 2002, ApJ, 577, 139

\bib Tozzi P. et al., 2001, ApJ, 562, 42

\bib Tran K.~H., Kelson D.~D., van Dokkum P., Franx M., Illingworth
G.~D., Magee D., 1999, ApJ, 522, 39

\bib van Dokkum P.~G., Franx M., Fabricant D., Illingworth G.~D.,
Kelson, D., 2000, ApJ, 541, 95

\bsp

\label{lastpage}

\end{document}

%% file: epsf_mn.tex
\newread\epsffilein    
\newif\ifepsffileok    
\newif\ifepsfbbfound   
\newif\ifepsfverbose   
\newdimen\epsfxsize    
\newdimen\epsfysize    
\newdimen\epsftsize    
\newdimen\epsfrsize    
\newdimen\epsftmp      
\newdimen\pspoints     
\def\epsfbox#1{\global\def\epsfllx{72}\global\def\epsflly{72}%
   \global\def\epsfurx{540}\global\def\epsfury{720}%
   \def\lbracket{[}\def\testit{#1}\ifx\testit\lbracket
   \let\next=\epsfgetlitbb\else\let\next=\epsfnormal\fi\next{#1}}%
\def\epsfgetlitbb#1#2 #3 #4 #5]#6{\epsfgrab #2 #3 #4 #5 .\\%
   \epsfsetgraph{#6}}%
\def\epsfnormal#1{\epsfgetbb{#1}\epsfsetgraph{#1}}%
\def\epsfgetbb#1{%
\openin\epsffilein=#1
\ifeof\epsffilein\errmessage{I couldn't open #1, will ignore it}\else
   {\epsffileoktrue \chardef\other=12
    \def\do##1{\catcode`##1=\other}\dospecials \catcode`\ =10
    \loop
       \read\epsffilein to \epsffileline
       \ifeof\epsffilein\epsffileokfalse\else
          \expandafter\epsfaux\epsffileline:. \\%
       \fi
   \ifepsffileok\repeat
   \ifepsfbbfound\else
    \ifepsfverbose\message{No bounding box comment in #1; using defaults}\fi\fi
   }\closein\epsffilein\fi}%
\def\epsfsetgraph#1{%
   \epsfrsize=\epsfury\pspoints
   \advance\epsfrsize by-\epsflly\pspoints
   \epsftsize=\epsfurx\pspoints
   \advance\epsftsize by-\epsfllx\pspoints
   \epsfxsize\epsfsize\epsftsize\epsfrsize
   \ifnum\epsfxsize=0 \ifnum\epsfysize=0
      \epsfxsize=\epsftsize \epsfysize=\epsfrsize
     \else\epsftmp=\epsftsize \divide\epsftmp\epsfrsize
       \epsfxsize=\epsfysize \multiply\epsfxsize\epsftmp
       \multiply\epsftmp\epsfrsize \advance\epsftsize-\epsftmp
       \epsftmp=\epsfysize
       \loop \advance\epsftsize\epsftsize \divide\epsftmp 2
       \ifnum\epsftmp>0
          \ifnum\epsftsize<\epsfrsize\else
             \advance\epsftsize-\epsfrsize \advance\epsfxsize\epsftmp \fi
       \repeat
     \fi
   \else\epsftmp=\epsfrsize \divide\epsftmp\epsftsize
     \epsfysize=\epsfxsize \multiply\epsfysize\epsftmp   
     \multiply\epsftmp\epsftsize \advance\epsfrsize-\epsftmp
     \epsftmp=\epsfxsize
     \loop \advance\epsfrsize\epsfrsize \divide\epsftmp 2
     \ifnum\epsftmp>0
        \ifnum\epsfrsize<\epsftsize\else
           \advance\epsfrsize-\epsftsize \advance\epsfysize\epsftmp \fi
     \repeat     
   \fi
   \ifepsfverbose\message{#1: width=\the\epsfxsize, height=\the\epsfysize}\fi
   \epsftmp=10\epsfxsize \divide\epsftmp\pspoints
   \newcount\figskipcount
      \message{#1 \the\epsfysize  }
   \vbox to\epsfysize{\vfil\hbox to\epsfxsize{%
      \includegraphics{#1}%
      \hfil}}%
\epsfxsize=0pt\epsfysize=0pt}%

{\catcode`\%=12 \global\let\epsfpercent=
\long\def\epsfaux#1#2:#3\\{\ifx#1\epsfpercent
   \def\testit{#2}\ifx\testit\epsfbblit
      \epsfgrab #3 . . . \\%
      \epsffileokfalse
      \global\epsfbbfoundtrue
   \fi\else\ifx#1\par\else\epsffileokfalse\fi\fi}%
\def\epsfgrab #1 #2 #3 #4 #5\\{%
   \global\def\epsfllx{#1}\ifx\epsfllx\empty
      \epsfgrab #2 #3 #4 #5 .\\\else
   \global\def\epsflly{#2}%
   \global\def\epsfurx{#3}\global\def\epsfury{#4}\fi}%
\def\epsfsize#1#2{\epsfxsize}